\documentclass[reprint,amsmath,amssymb,aps,pra,floatfix]{revtex4-2}

\usepackage[T1]{fontenc}
\usepackage{lmodern}
\usepackage{microtype}
\usepackage{amsthm}
\usepackage{braket}
\usepackage{float}
\usepackage{newtxtext,newtxmath} % optional font upgrade
\usepackage[version=4]{mhchem}

\usepackage[
    colorlinks=true,
    citecolor=blue,   % Color for citations
    urlcolor=blue,    % Color for URLs
    linkcolor=blue    % Color for internal document links (like eq or fig refs)
]{hyperref}

\usepackage{orcidlink}
\usepackage{braket}
\usepackage{graphicx}
\usepackage{dcolumn}
\usepackage{bm}
\usepackage{siunitx}
\usepackage{amsthm}
\usepackage{booktabs}
\usepackage{nicematrix}

\usepackage{tikz}
\usetikzlibrary{arrows.meta, decorations.pathreplacing}

\usepackage[mathlines]{lineno}

\DeclareSIUnit{\au}{a.u.}

\begin{document}

\title{Unified Strong-Field Dynamics Simulations from Atoms to Heterostructures}

\author{Zakaria Dahbi \orcidlink{0000-0001-9933-2184}}
\email{zakaria.dahbi@kcl.ac.uk}
\author{Amelle Za\"ir \orcidlink{0000-0003-1687-5453}}%
\affiliation{Attosecond Quantum Physics Laboratory, Department of Physics, King’s College London, Strand Campus, WC2R 2LS, UK}

\date{\today}

\begin{abstract}
We present \textsc{TDSE-Z}, a high-performance open-source framework for strong-field quantum dynamics in atomic, molecular, and semiconductor effective-mass systems. The core engine implements a weak-form Galerkin discretisation of the Hermitian BenDaniel-Duke operator, $\hat{T}_{\mathrm{BDD}} = -\frac{1}{2}\nabla\cdot(m^{-1}(\mathbf{r})\nabla)$, on geometry-adapted B-spline meshes, supporting arbitrary potentials and customisable laser configurations in one to three dimensions. 
We validate the static position-dependent-mass (PDM) eigensolver through two stringent benchmarks: a comparison to the analytical Quesne PDM model and a $\text{GaAs/Al}_{0.3}\text{Ga}_{0.7}\text{As}$ double quantum well, where the exponential decay of computed tunnel splittings follows Wentzel–Kramers–Brillouin (WKB) theory at the sub-percent level. We further demonstrate the time-propagation engine on constant-mass systems, accurately reproducing high-harmonic generation (HHG) spectra in atomic benchmarks and confirming the importance of dimensionality in fully capturing the strong light-matter interaction. Our implementation demonstrates robust strong-scaling efficiency, maintaining performance across hundreds of CPU cores. While the static eigensolver currently supports optional GPU offloading, the time-propagation engine is CPU-optimised, providing a modular architecture for future expansion toward exascale quantum dynamics.
\end{abstract}

\maketitle

\section{Introduction}
\label{sec:intro}
The interaction of intense, ultrashort laser pulses with matter drives some of the most extreme nonlinear optical phenomena known, including HHG, above-threshold ionisation, and attosecond pulse production \cite{krausz2009attosecond, corkum1993plasma}. These processes are commonly modelled using the single-active-electron (SAE) time-dependent Schrödinger equation (TDSE). In the presence of a spatially varying effective mass, such as in heterostructured systems, the Hamiltonian couples the wavefunction to a potentially discontinuous mass profile via the BenDaniel–Duke (BDD) kinetic operator \cite{bendaniel1966space}. For homogeneous atomic gases, the standard Laplacian operator $-\tfrac{1}{2}\nabla^2$ $(m^*\equiv 1$) is exact. For condensed-matter and low-dimensional systems, the BDD operator $-\tfrac{1}{2}\nabla\cdot(m^{*-1}\nabla)$ keeps the Hamiltonian self-adjoint and preserves the $(1/m^*)\nabla\psi$ probability flux at material interfaces. Numerically, this poses a distinct challenge; any scheme that does not enforce the BDD weak form introduces spurious interface reflections that can manifest as spurious physical effects \cite{zhu1983interface}.

This technical challenge is not merely academic: it blocks the advancement of a neat description quantum technologies. Semiconductor heterostructures and quantum dots represent leading solid-state platforms for ultrafast quantum information processing and quantum photonic devices  \cite{hennessy2007quantum,khitrova2006vacuum}. However, the ultrafast electron dynamics that govern their operational limits arise directly from the spatially varying effective mass and sharp material interfaces of these engineered structures. Accurately simulating the strong-field response of these systems—a prerequisite for designing next-generation attosecond optoelectronics—thus requires a solver that rigorously enforces BDD boundary conditions while simultaneously supporting robust time-dependent propagation. Recent solid-state HHG experiments further underscore this need in semiconductor heterostructures and doped quantum wells \cite{ghimire2011observation, vampa2015semiclassical, liu2017high}, while parallel advances in molecular attosecond spectroscopy continue to demand high-fidelity simulations to decode the complex electronic and nuclear dynamics observed in experimental measurements \cite{zhao2026floqueteng, zou20262d}. 

A wide and growing application landscape now demands the BDD operator at the strong-field level, including the harmonic response of semiconductor heterostructures, electron-hole dynamics in Moir\'e heterostructures, and band-resolved intra-/inter-band dynamics. However, a robust framework capable of efficiently handling the BenDaniel–Duke (BDD) operator for these structures has been notably lacking in current strong-field solvers. B-spline basis sets have become a standard for TDSE simulations in atomic and molecular physics \cite{cormier1997above, bachau2001applications} due to their high-order convergence and flexibility when combined with non-uniform knot distributions; such representations enable efficient resolution of both short-range structure and long-range continuum oscillations while maintaining systematic convergence. Existing public software occupies adjacent but disjoint specialisations: B-spline atomic codes (e.g., QPROP \cite{bauer2006qprop}) assume constant mass; Cartesian FFT split-operator codes (e.g., PCTDSE \cite{fu2017pctdse}, 3D-GTDSE \cite{peng20253d}) likewise assume constant mass in the kinetic term; and spherical-coordinate codes (e.g., SCID-TDSE \cite{patchkovskii2016simple}) support central potentials without generalising the mass profile. While each is highly effective in its own right, none simultaneously provides the flexibility of B-splines with the BDD formalism required for spatially varying effective-mass systems.

In this paper, we close this gap by presenting \textsc{TDSE-Z}, a unified framework built upon a weak-form Galerkin discretisation of the BDD kinetic operator on geometry-adapted B-spline meshes. We validate the framework across two distinct tiers. First, a machine-precision match to the analytical Quesne Position-Dependent Mass (PDM) benchmark establishes the correctness of the weak-form BDD eigensolver. Second, we verify the constant-mass reduction to the canonical atomic TDSE against the hydrogenic Rydberg series and the HHG cutoff law. For further consistency checks, we performed a stringent validation of the time-propagation engine against analytic harmonic oscillator predictions (see appendix). Beyond this validation, we apply the framework to a realistic $\text{GaAs/Al}_{0.3}\text{Ga}_{0.7}\text{As}$ double quantum well, where computed static tunnel splittings replicate the WKB exponential decay over more than two orders of magnitude. Built upon the PETSc \cite{balay2025petsc, abhyankar2018petsc}, SLEPc \cite{hernandez2005slepc}, and PetIGA \cite{dalcin2016petiga} ecosystem, the framework demonstrates robust strong-scaling efficiency, exceeding $80\%$ up to $128$ cores. The architecture is designed for heterogeneous computing: the Time-Independent Schr\"odinger equation (TISE) eigensolver is fully GPU-accelerated for rapid initial-state preparation, and the time-propagation engine leverages highly optimised CPU-based MPI parallelism. To maximise transparency of our solver
 and its reproducibility, \textsc{TDSE-Z} is available for non-commercial academic use, providing the Atomic-Molecule-Optics (AMO), condensed-matter, and computational physics communities with a production-ready platform to explore laser-driven dynamics across diverse spatial and mass scales.

\section{BenDaniel-Duke TDSE}
\label{sec:bdd}

Within the SAE approximation, the wavefunction $\Psi(\mathbf{r},t)$ satisfies the time-dependent Schrödinger equation
\begin{equation}
i\partial_t \Psi(\mathbf{r},t)
=
\left[
-\tfrac{1}{2}\nabla\cdot\left(m^{-1}(\mathbf{r})\nabla\right)
+ V(\mathbf{r})
+ V_{\mathrm{L}}(\mathbf{r},t)
\right]\Psi(\mathbf{r},t),
\label{eq:tdse}
\end{equation}
where $m(\mathbf{r})$ denotes a spatially varying effective mass, $V(\mathbf{r})$ is the static potential, and $V_{\mathrm{L}}$ describes the external laser field.

\subsection{BDD kinetic operator}

In systems with spatially varying effective mass, such as semiconductor heterostructures, the standard kinetic operator $-\tfrac{1}{2}\nabla^2$ fails to conserve the probability current at material interfaces where the mass $m(\mathbf{r})$ is discontinuous. We address this using the BenDaniel-Duke kinetic operator \cite{bendaniel1966space},
\begin{equation}
\hat{T}_{\mathrm{BDD}} = -\tfrac{1}{2}\nabla\cdot\left(\frac{1}{m(\mathbf{r})}\nabla\right),
\end{equation}
which recovers the standard Laplacian in the constant-mass limit ($m(\mathbf{r}) \equiv 1$). This operator is self-adjoint with respect to the standard $L^2(\Omega)$ inner product, defined for any two functions $\phi, \psi \in L^2(\Omega)$ as $\langle \phi, \psi \rangle = \int_{\Omega} \phi^*(\mathbf{r}) \psi(\mathbf{r}) \, d\mathbf{r}$. This property is fundamental: it ensures the continuity of both the wavefunction $\psi(\mathbf{r})$ and the probability current $\mathbf{J} \propto \frac{1}{m(\mathbf{r})} \nabla \psi(\mathbf{r})$ across interfaces without requiring additional explicit interface boundary conditions. In our framework, these requirements are naturally satisfied by the weak-form Galerkin discretisation, which effectively resolves the mass-gradient singularities arising in heterogeneous nanostructures.

\subsection{Laser--matter coupling}
In the length gauge, the electron couples to the external laser field via the multiplicative potential
\begin{equation}
V_{\rm L}(\mathbf{r},t) = \mathbf{E}(t)\cdot\mathbf{r},
\label{eq:laser}
\end{equation}
where $\mathbf{E}(t)$ is the electric field of the pulse. This is the primary and default gauge implemented in TDSE-Z for time propagation. The length gauge Hamiltonian is
\begin{equation}
\hat{H}(t) = \hat{T}_{\rm BDD} + V(\mathbf{r}) + \mathbf{E}(t)\cdot\mathbf{r},
\end{equation}
with the electric field $\mathbf{E}(t)$ specified as a runtime evaluable string. The BDD velocity operator becomes 
\begin{equation}
\hat{\mathbf{v}}_{\rm BDD} = -i\, m^{-1}(\mathbf{r})\nabla - \frac{i}{2}\nabla m^{-1}(\mathbf{r}),
\end{equation}
which is Hermitian. As a litmus test of the quality of the discretisation, the code verifies length–velocity equivalence via the commutation relation 
\begin{equation}
v^{(\alpha)}_{mn} \stackrel{!}{=} i(\varepsilon_n-\varepsilon_m)d^{(\alpha)}_{mn},
\end{equation}
for every non-degenerate state pair $(m,n)$ of energies $(\varepsilon_m, \varepsilon_n)$, where $\alpha \in \{x, y, z\}$ denotes the spatial component. This is necessary to confirm gauge invariance of the stationary dipole/velocity matrices before initiating any quantum simulation.

\subsection{Weak form}
The BDD operator is treated in weak form. Multiplying the TDSE by a test function $\phi(\mathbf{r})\in H_0^1(\Omega)$, integrating over $\Omega$, and applying Green’s theorem yields
\begin{align}
\bigl\langle \phi, i\partial_t \Psi \bigr\rangle
&= \frac{1}{2}\bigl\langle \nabla\phi,\, m^{-1}(\mathbf{r})\nabla\Psi \bigr\rangle_{\Omega} + \bigl\langle \phi,\, V(\mathbf{r})\Psi \bigr\rangle_{\Omega} \nonumber\\
&\quad+ \bigl\langle \phi,\, V_{\rm L}(\mathbf{r},t)\Psi \bigr\rangle_{\Omega} - \frac{1}{2}\int_{\partial\Omega} \overline{\phi}\, m^{-1}(\mathbf{r})\frac{\partial\Psi}{\partial n}\,dS.
\label{eq:weak}
\end{align}
where $dS$ is the surface area element, and $\frac{\partial\Psi}{\partial n} = \mathbf{n} \cdot \nabla\Psi$ denotes the outward normal derivative at the boundary $\partial\Omega$. The boundary term arises from the identity
\begin{align}
\int_{\Omega} \overline{\phi}\,\nabla\cdot(m^{-1}\nabla\Psi)\,d\mathbf{r} 
&= -\int_{\Omega} (\nabla\overline{\phi})\cdot (m^{-1}\nabla\Psi)\,d\mathbf{r} \nonumber \\
&+ \oint_{\partial\Omega} \overline{\phi}\, m^{-1}\frac{\partial\Psi}{\partial n}\,dS.
\end{align}
The surface integral is rigorously eliminated under general conditions. It vanishes identically if Neumann conditions $\partial_n \Psi|_{\partial\Omega}=0$ are imposed, or if Dirichlet conditions $\bar{\phi}|_{\partial\Omega}=0$ are imposed. These conditions cancel the surface term and make the interior weak formulation fully general and independent of the specific boundary treatment employed. The resulting weak form becomes
\begin{align}
\langle \phi, i\partial_t \Psi\rangle = \frac{1}{2} \int_{\Omega} (\nabla\overline{\phi})\cdot(m^{-1}(\mathbf{r})\nabla\Psi)\,d\mathbf{r} + \int_{\Omega} \overline{\phi}\,(V + V_{\rm L})\Psi\,d\mathbf{r},
\end{align}
where the first term on the right-hand side demonstrates the self-adjointness of the BDD kinetic energy. As the mass is always positive ($m(\mathbf{r})>0$), this form is bounded and positive-definite, and involves only first-order derivatives. The three-dimensional basis is constructed as a tensor product of univariate splines. Let $\Xi^{(x)}$, $\Xi^{(y)}$, $\Xi^{(z)}$ be the knot vectors in the $x, y, z$ directions, with $n_x, n_y, n_z$ basis functions, respectively. A multi-index $I = (i_x,i_y,i_z)$ labels the 3D basis function
\begin{equation}
B_I(\mathbf{r}) = N_{i_x,p}(\xi_x)\, N_{i_y,p}(\xi_y)\, N_{i_z,p}(\xi_z),
\label{eq:tensor3d}
\end{equation}
where $\xi_x,\xi_y,\xi_z$ are the local coordinates. We enforce homogeneous Dirichlet conditions using open knot vectors in each direction: the first and last knots are repeated $p+1$ times, making the basis interpolatory at the boundaries. The boundary basis functions ($i_\alpha=1$ and $i_\alpha=n_\alpha$) take non-zero values only at the domain edges; constraining their coefficients to zero ensures that all basis functions lie strictly in $H_0^1(\Omega)$.

The knot vectors must be engineered to resolve the Coulomb cusp, bound-state oscillations, potential geometry, and long-range continuum. The finite expansion of the wavefunction is
\begin{equation}
\Psi(\mathbf{r},t) = \sum_I c_I(t)\, B_I(\mathbf{r}),
\label{eq:expand}
\end{equation}
where $c_I(t)$ are time-dependent coefficients. Substituting this into the weak form~\eqref{eq:weak} (after removal of the boundary term) and testing with each $B_J$ yields
\begin{align}
i\sum_I \langle B_J, B_I\rangle\,\dot{c}_I(t)
&= \frac{1}{2}\sum_I \langle \nabla B_J,\, m^{-1}\nabla B_I\rangle\,c_I(t) \nonumber\\
&+ \sum_I \langle B_J,\, (V + V_{\rm L}) B_I\rangle\,c_I(t).
\label{eq:semi}
\end{align}
By introducing the overlap matrix $\mathbf{M}$ and the time-dependent Hamiltonian matrix $\mathbf{H}(t)$, defined as
\begin{align}
M_{JI} &= \langle B_J, B_I\rangle, \nonumber \\
H_{JI}(t) &= \frac{1}{2}\langle \nabla B_J,\, m^{-1}\nabla B_I\rangle + \langle B_J,\, (V + V_{\rm L}) B_I\rangle,
\end{align}
substituting the expansion into the weak form leads to a system of coupled first-order ordinary differential equations, which can be written in matrix form as
\begin{equation}
i\,\mathbf{M}\,\dot{\mathbf{c}}(t) = \mathbf{H}(t)\,\mathbf{c}(t).
\label{eq:matrixODE}
\end{equation}
Because each $B_I$ has compact support, both $\mathbf{M}$ and $\mathbf{H}$ are sparse and banded. All inner products are evaluated element-wise using Gaussian quadrature on the knot spans. At $t=0$, the laser is switched off ($V_{\rm L}=0$) and the time-dependence of the stationary states separates as $\Psi(\mathbf{r},t)=\psi_n(\mathbf{r})\,e^{-i \varepsilon_n t}$. Inserting the B-spline expansion $\psi_n(\mathbf{r})=\sum_I c_{I}^{(n)}B_I(\mathbf{r})$ into the weak form~\eqref{eq:weak} yields the generalised eigenvalue problem
\begin{equation}
\mathbf{H}^{(0)}\,\mathbf{c}_n = \varepsilon_n \,\mathbf{M}\,\mathbf{c}_n,
\label{eq:gen_ev}
\end{equation}
where $\mathbf{H}^{(0)}$ is the time-independent Hamiltonian matrix. Equation~\eqref{eq:gen_ev} is solved with the SLEPc library using a shift-and-invert Krylov–Schur method \cite{hernandez2005slepc}. Inner linear systems are handled by GMRES with block-Jacobi or algebraic multigrid preconditioning, depending on the problem size \cite{balay2025petsc,abhyankar2018petsc}. The obtained eigenpairs $(E_n,\mathbf{c}_n)$ provide the bound-state energies and the coefficient vectors that serve as the initial state for time propagation.

\section{Knot engineering}
\begin{figure}[ht!]
 \centering
 \includegraphics[width=0.9\linewidth]{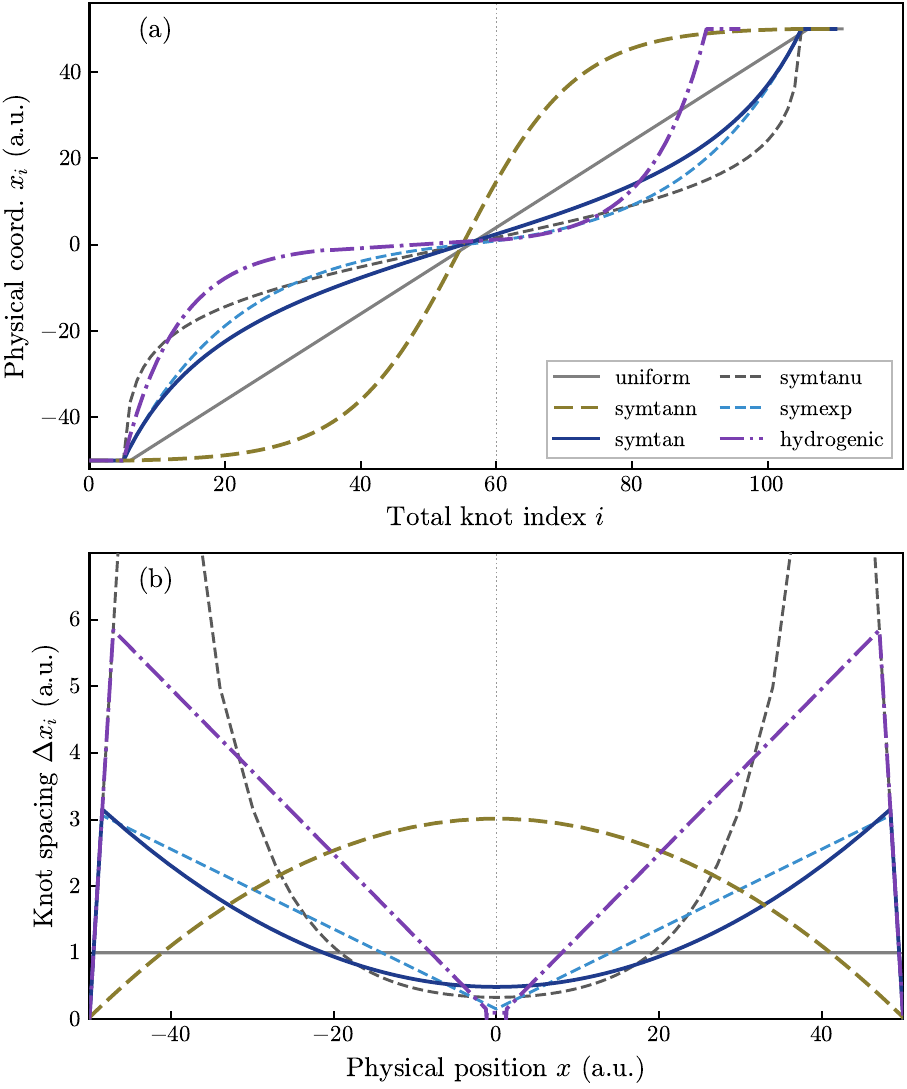}
    \caption{Illustration of knot distributions and local grid resolution for the six coordinate mappings.
    (a) Physical knot positions $x_i$ vs.\ index $i$ over $[-50,50]$~a.u.\ ($N_{\mathrm{interior}}=99$, $p=5$). Clustering near the origin is evident for all non-uniform schemes.
    (b) Grid spacing $\Delta x_i = x_{i+1}-x_i$. The symtan, symexp, and hydrogenic mappings achieve sub-atomic resolution ($<0.2$~a.u.) near the Coulomb cusp without increasing the global basis size.}
 \label{fig:knotviz}
\end{figure}

The configuration of the spatial knot sequence is the primary determinant of accuracy and numerical stability in B-spline discretisations. For intense laser–atom interactions, the mesh must resolve distinct length scales: the steep nuclear Coulomb cusp at the origin, the oscillatory bound-state wavefunctions in the core, and the long-wavelength continuum wavepackets that propagate over hundreds of atomic units. While a uniform mesh resolves short-range features, it wastes substantial degrees of freedom (DoFs) in the outer asymptotic regions, becoming computationally prohibitive for large spatial grids. Conversely, purely exponential meshes compress spacing near the origin but fail to resolve the oscillatory structure of higher-lying bound states. To address these competing requirements, the present version of TDSE-Z implements six distinct non-uniform knot grading strategies, summarised in Table \ref{tab:gradings}. Each knot sequence optimises the localised spatial representation for specific physical regimes. A pictorial representation of these knot sequences is shown in Fig. \ref{fig:knotviz}.

\begin{table}[ht!]
\caption{Implemented coordinate grading functions mapping a uniform parameter $t_i \in [-1,1]$ (or index $i$) to physical grid coordinates $x_i$ or $r_i$ over the domain $[L_{\mathrm{min}}, L_{\mathrm{max}}]$. Clustering and transition properties are regulated by parameters $\alpha, \beta, \gamma > 0$, while $N_{\mathrm{elem}}$ denotes the total number of intervals or grid points. For the \texttt{hydrogenic} scheme, $n_{\mathrm{lin}}$ is the number of linear inner intervals of step size $r_1$, $r_{\mathrm{cross}} = n_{\mathrm{lin}} r_1$ is the crossover radius separating the inner linear region from the outer domain, and $n_{\mathrm{exp}}$ is the number of exponentially expanding intervals extending up to $L_{\max}$.}
\label{tab:gradings}
\footnotesize
\centering
\setlength{\tabcolsep}{4pt}
\begin{tabular}{ll}
\toprule
\textbf{Name} & \textbf{Coordinate Mapping Formula} \\
\midrule
\texttt{Uniform} & $x_i = -L_{\max} + i \frac{2L_{\max}}{N_{\mathrm{elem}}}$ \\[6pt]
\texttt{symexp} & $x_i = \operatorname{sgn}(t_i)L_{\max} \frac{e^{\alpha|t_i|}-1}{e^{\alpha}-1}$ \\[8pt]
\texttt{symtan} & $x_i = L_{\max}\frac{\tan(\alpha_{\mathrm{s}}t_i)}{\tan\alpha_{\mathrm{s}}},\ \alpha_{\mathrm{s}}=\frac{\pi}{2}\frac{\alpha}{\alpha+1}$ \\[8pt]
\texttt{symtann} & $x_i = L_{\min}+\frac{L_{\max}-L_{\min}}{2}\left(1+\frac{\tanh(\beta t_i)}{\tanh\beta}\right)$ \\[8pt]
\texttt{symtanu} & $x_i = L_{\min}+\frac{L_{\max}-L_{\min}}{2}\left(1+\frac{\operatorname{atanh}(t_i\tanh\beta)}{\beta}\right)$ \\[6pt]
\texttt{hydrogenic} & $\begin{cases} 
r_i = i\,r_1 & (0 \le i \le n_{\mathrm{lin}}) \\[4pt] 
r_i = r_{\mathrm{cross}} \exp\!\left(\frac{(i - n_{\mathrm{lin}})\ln(L_{\max}/r_{\mathrm{cross}})}{n_{\mathrm{exp}}}\right) & (n_{\mathrm{lin}} < i \le n_{\mathrm{lin}} + n_{\mathrm{exp}}) 
\end{cases}$ \\
\bottomrule
\end{tabular}
\end{table}

The adaptive knot sequences place grid points where the wavefunction varies most rapidly—near the Coulomb singularity and in the oscillatory bound-state region—while maintaining a sparse distribution in the long-range continuum. This non-uniform allocation minimises the total number of DoFs needed for a target accuracy and avoids numerical artefacts arising from uniform finite-difference grids. This has the advantage of cutting the computational cost of eigensolver and time propagator by over an order of magnitude compared to uniform meshes of comparable resolution.

\section{Physical quantities}
\subsection{Dipoles}
We compute the length-gauge dipole moment directly from the time-dependent coefficients $\mathbf{c}(t)$:
\begin{equation}
\mathbf{d}(t) = \langle \Psi(t) | \mathbf{r} | \Psi(t) \rangle = \mathbf{c}^\dagger(t) \mathbf{D} \mathbf{c}(t),
\end{equation}
where $\mathbf{D}_{\alpha,IJ} = \langle B_I | \hat{r}_\alpha | B_J \rangle$ is the precomputed dipole matrix. We evaluate the dipole acceleration in the length gauge using Ehrenfest's theorem:

We evaluate the dipole acceleration directly from Ehrenfest’s theorem:
\begin{equation}
\mathbf a(t)
=
-\left\langle\Psi(t)\left|
m^{-1}(\mathbf r)\bigl[\nabla V(\mathbf r)+\mathbf E(t)\bigr]
\right|\Psi(t)\right\rangle
+\mathbf a_{\rm Q}(t),
\label{eq:pdm_acceleration}
\end{equation}
where the Cartesian components of the position-dependent-mass kinetic contribution are
\begin{widetext}
\begin{equation}
a_{{\rm Q},\alpha}(t)
= \frac{i}{4}\sum_k
\left\langle\Psi(t)\left|
\left[
\hat p_k m^{-1}(\mathbf r)\hat p_k,\,
\hat p_\alpha m^{-1}(\mathbf r)
+m^{-1}(\mathbf r)\hat p_\alpha
\right]
\right|\Psi(t)\right\rangle.
\label{eq:aQ_definition}
\end{equation}
\end{widetext}
Here, $\hat p_\alpha=-i\partial_\alpha$, and $k$ runs over the Cartesian directions. The term $\mathbf a_{\rm Q}(t)$ captures the kinetic effect of the spatially varying effective mass and vanishes in the constant-mass limit. Evaluating Eq.~\eqref{eq:pdm_acceleration} directly avoids the numerical noise introduced by differentiating $\mathbf d(t)$ twice. 
\\

We compute both observables concurrently during time propagation, providing on-the-fly data for HHG analysis. To suppress unphysical edge artefacts resulting from the finite duration of the simulation window, an apodisation function $W(t)$—such as a Kaiser window—is applied to the signals. The resulting HHG spectral intensity $S(\omega)$ follows from the windowed Fourier transform:
\begin{align}
S(\omega) &= \left| \int_{-\infty}^{\infty} a(t) W(t)\, e^{i\omega t}\,{\rm d}t \right|^2 \approx \omega^4 \left| \int_{-\infty}^{\infty} d(t) W(t)\, e^{i\omega t}\,{\rm d}t \right|^2,
\end{align}
demonstrating the structural connection between the acceleration and length formulations up to an $\omega^4$ scaling factor.

\subsection{Currents and coherence dynamics}
\label{sec:currents_coherence}
The physical quantity required for the harmonic spectra is the time derivative of the dipole moment, which in the length gauge is given by the expectation value of the velocity operator. For the BenDaniel–Duke Hamiltonian, the Hermitian velocity operator is
\begin{equation}
 \hat v_\alpha = \frac{1}{2}\left( \hat p_\alpha\frac{1}{m(\mathbf{r})} + \frac{1}{m(\mathbf{r})}\hat p_\alpha \right), \qquad \hat p_\alpha = -i\partial_\alpha,
 \label{eq:velocity_op}
\end{equation}
which reduces to standard momentum when the effective mass is constant. The total dipole velocity is defined as
\begin{equation}
 J_\alpha^{\rm tot}(t) \equiv \langle\psi(t)|\hat v_\alpha|\psi(t)\rangle = \frac{d}{dt}\langle\psi(t)|\hat r_\alpha|\psi(t)\rangle.
\end{equation}
Expanding the wavefunction in the field-free eigenstates, $\psi(\mathbf{r},t) = \sum_n c_n(t)\varphi_n(\mathbf{r})$, and using the precomputed velocity matrix elements $v_{\alpha,mn} = \langle\varphi_m|\hat v_\alpha|\varphi_n\rangle$, the total dipole velocity decomposes into three contributions:
\begin{align}
J_\alpha^{\rm tot}(t) &= J_\alpha^{\rm intra}(t) + J_\alpha^{\rm inter}(t) + J_\alpha^{\rm bc}(t), \label{eq:Jdecomp} \\
J_\alpha^{\rm intra}(t) &= \sum_n |c_n(t)|^2\,v_{\alpha,nn}, \\
 J_\alpha^{\rm inter}(t) &= \sum_{m\neq n} c_m^*(t)c_n(t)\, v_{\alpha,mn}, \\
J_\alpha^{\rm bc}(t) &= J_\alpha^{\rm tot}(t) - J_\alpha^{\rm intra}(t) - J_\alpha^{\rm inter}(t). \label{eq:Jbc}
\end{align}
The diagonal ($m=n$) term $J_\alpha^{\rm intra}$ is the intraband current (bound-state coherence contribution); the off-diagonal term $J_\alpha^{\rm inter}$ is driven by interband coherences $c_m^*c_n$. The remainder $J_\alpha^{\rm bc}$ accounts for continuum components not represented by the finite set of bound states, which is essential when ionisation occurs. For time-reversal invariant systems without magnetic fields, eigenstates may be chosen real, $\varphi_n\in\mathbb{R}$. In that case, the velocity matrix elements are purely imaginary, $v_{\alpha,nn}=0$, and $J_\alpha^{\rm intra}$ vanishes identically. The entire bound-state contribution to the dipole velocity is then interband, driven by the off-diagonal elements of the single-particle density matrix $\varrho_{mn}=c_m^*c_n$. The coefficients $c_n(t)$ obey the equations of motion $i\dot c_m = \varepsilon_m c_m + \mathbf{E}(t)\cdot\sum_n \mathbf{d}_{mn}c_n$, leading to the optical Bloch equation
\begin{equation}
 i\,\partial_t\varrho_{mn} = (\varepsilon_n-\varepsilon_m)\varrho_{mn} + \mathbf{E}(t)\cdot\sum_k\bigl( \mathbf{d}_{nk}\varrho_{mk} - \mathbf{d}_{km} \varrho_{kn} \bigr),
 \label{eq:bloch}
\end{equation}
which governs the coherent dynamics. The interband current $J_\alpha^{\rm inter}$ is therefore directly proportional to the instantaneous quantum coherence between eigenstates, and the velocity matrix elements satisfy the length–velocity relation $v_{\alpha,mn} = i (\varepsilon_n-\varepsilon_m) d_{\alpha,mn}$.

\subsection{Wavefunction realignment and commutator verification}

Numerical eigenstates obtained from SLEPc \cite{hernandez2005slepc} may carry an arbitrary complex global phase. For visualisation and symmetry analysis, the code can realign each eigenvector so that it becomes purely real. This is achieved by computing the $\mathbf{M}$-weighted overlap $\langle \phi_n^*| \mathbf{M} | \phi_n^* \rangle$ (where $\mathbf{M}$ is the overlap matrix) to determine the global phase, then rotating the vector by $\exp(-i\,\text{arg}(\langle \phi_n^*|\mathbf{M}|\phi_n^*\rangle)/2)$. The resulting real vector is then renormalised to unity in the $\mathbf{M}$-norm. Symmetry and conservation laws are tested by evaluating commutators at the discrete matrix level. The angular momentum commutator $[\mathbf{H}, \hat{L}_z]$ is assembled via sparse matrix-matrix multiplication,
\begin{equation}
\mathbf{C} = \mathbf{H}\mathbf{M}^{-1}\mathbf{L}_z - \mathbf{L}_z \mathbf{M}^{-1} \mathbf{H},
\end{equation}
and its Frobenius norm $\|\mathbf{C}\|_{\mathrm{F}}$ is reported. In a spherically symmetric potential, $\|\mathbf{C}\|_{\mathrm{F}}$ should vanish to machine precision, a property verified empirically during the eigensolution loop.

 \section{Numerical Implementation}
 \subsection{Parallel assembly}
We assemble the element-wise matrices in parallel using PetIGA, which distributes elements across MPI ranks and handles the communication of shared DoFs on partition boundaries. The global sparse matrices are stored in PETSc's \texttt{Mat} format, enabling direct access to PETSc's KSP (Krylov subspace methods), PC (preconditioners), and EPS (eigensolver) interfaces. The code uses \texttt{MatType MATMPIAIJ} for distributed matrices and \texttt{MatType MATAIJCUSPARSE} for GPU-accelerated assembly. Assembly time scales linearly with the number of elements and shows near-ideal strong scaling. For each element $e$ with local coordinates $\boldsymbol{\xi}\in[-1,1]^d$, the quadrature points $\boldsymbol{\xi}_q$ and weights $w_q$ are provided by PetIGA. The Jacobian $J_e(\boldsymbol{\xi}_q)$ maps to physical coordinates $\mathbf{r}_q$. At each quadrature point, the basis functions $N_a(\boldsymbol{\xi}_q)$ and their gradients $\nabla_{\xi}N_a(\boldsymbol{\xi}_q)$ are evaluated. Physical gradients are obtained via $\nabla_{\mathbf{r}}N_a = \mathbf{J}_e^{-T}\nabla_{\boldsymbol{\xi}}N_a$. The elemental contributions to the global system—where indices $a, b$ run over local basis functions, $q$ over quadrature points, and $J_q = \det(\mathbf{J}_e(\boldsymbol{\xi}_q))$ is the determinant of the Jacobian matrix—are assembled as follows:

\begin{itemize}
\item \textbf{Overlap matrix:}
\begin{equation}
\mathbf{M}^e_{ab} = \sum_q w_q J_q\, N_a(\boldsymbol{\xi}_q)\, N_b(\boldsymbol{\xi}_q).
\label{eq:mass_elem}
\end{equation}

\item \textbf{BDD kinetic energy matrix:}
\begin{equation}
\mathbf{T}^e_{ab} = \frac{1}{2} \sum_q w_q J_q\, m^{-1}(\mathbf{r}_q)\, \bigl(\nabla_{\mathbf{r}} N_a(\boldsymbol{\xi}_q)\bigr)\!\cdot\!\bigl(\nabla_{\mathbf{r}} N_b(\boldsymbol{\xi}_q)\bigr).
\label{eq:bdd_elem}
\end{equation}

This is the discrete analogue of the weak form integral $\frac{1}{2}\langle\nabla v, m^{-1}\nabla\Psi\rangle$. For position-dependent effective mass, the quadrature evaluates $m^{-1}(\mathbf{r}_q)$ pointwise, ensuring mass discontinuities are captured automatically when the mesh resolves the interface. The element matrix is symmetric positive-definite. 

\item \textbf{Potential and Dipole matrices:}
\begin{align}
\mathbf{V}^e_{ab} &= \sum_q w_q J_q\, V(\mathbf{r}_q)N_a(\boldsymbol{\xi}_q)\, N_b(\boldsymbol{\xi}_q),\\ 
\mathbf{D}^e_{ab} &= \sum_q w_q J_q\, \mathbf{r}_q\, N_a(\boldsymbol{\xi}_q)\, N_b(\boldsymbol{\xi}_q).
\end{align}

\item \textbf{Complex Absorbing Potential (CAP) matrix:}
\begin{equation}
\mathbf{W}^e_{ab} = -i \sum_q w_q J_q\, \eta(\mathbf{r}_q)\, N_a(\boldsymbol{\xi}_q)\, N_b(\boldsymbol{\xi}_q),
\end{equation}
where $\eta(\mathbf{r})$ is a smooth mask function, non-zero only in the boundary layer, implemented via the Manolopoulos-type CAP \cite{manolopoulos2002derivation}.

\item \textbf{Velocity matrix:}
The BDD velocity operator is assembled as
\begin{equation}
\mathbf{Vel}^e_{\alpha,ab} = -\frac{i}{2} \sum_q w_q J_q\, m^{-1}(\mathbf{r}_q)\, \Bigl( N_a\,\partial_\alpha N_b - \partial_\alpha N_a\, N_b \Bigr).
\label{eq:vel_elem}
\end{equation}
The full matrix $\mathbf{Vel}_\alpha$ is Hermitian and, after projection onto eigenstates, yields the matrix elements $v_{\alpha,mn}$ used in the current decomposition.

\item \textbf{Angular momentum operator:}
For two-dimensional systems, $\hat L_z = -i(x\partial_y - y\partial_x)$ is assembled via
\begin{equation}
\mathbf{L}^e_{ab} = -i \sum_q w_q J_q\, N_a(\boldsymbol{\xi}_q)\, \bigl( x_q\,\partial_y N_b(\boldsymbol{\xi}_q) - y_q\,\partial_x N_b(\boldsymbol{\xi}_q) \bigr).
\label{eq:Lz_elem}
\end{equation}
\end{itemize}
Global matrices are formed by summing elemental contributions. To maximise efficiency and reduce bandwidth,  we introduce a new technique (\texttt{TDSEZCompOperators}) that assembles a large collection of global matrices in a single, fused pass. 

\subsection{Time propagation}
\label{sec:time_prop}
We propagate the wavefunction using  Crank-Nicolson scheme, which can be viewed as the average of implicit and explicit Euler steps. This method is second-order accurate, unconditionally stable, and norm-conserving for hermitian Hamiltonians $\mathbf{H}(t)$:
\begin{equation}
\bigl( \mathbf{M} + \tfrac{i\Delta t}{2} \mathbf{H}^{n+1/2} \bigr) \mathbf{c}^{n+1} = \bigl( \mathbf{M} - \tfrac{i\Delta t}{2} \mathbf{H}^{n+1/2} \bigr) \mathbf{c}^n,
\label{eq:cn_system}
\end{equation}
where $\mathbf{H}^{n+1/2} = \frac{1}{2}(\mathbf{H}^{n+1}+\mathbf{H}^n)$. The linear system is solved with GMRES preconditioned by additive Schwarz (\texttt{PCASM}) with overlap 1; each subdomain uses ILU(0) factorisation and a direct solver (\texttt{KSPPREONLY}). Solver tolerances are set to $\texttt{rtol}=10^{-12}$ and $\texttt{atol}$ at machine precision. For large-scale problems, the preconditioner may be switched at runtime to algebraic multigrid (\texttt{PCGAMG}). The scheme inherently conserves the discrete $M$-norm, $\langle \mathbf{c}^{n+1} | \mathbf{M} | \mathbf{c}^{n+1} \rangle = \langle \mathbf{c}^n | \mathbf{M} | \mathbf{c}^n \rangle$, to solver tolerance. Benchmarks against exact analytic solutions of the quantum harmonic oscillator confirm that the energy error remains at machine precision, validating the long-term norm conservation of the scheme. For strong-field intensities exceeding $E_0 \approx 0.5$~a.u., we adopt smaller time steps ($\Delta t < 0.01$~a.u.) to accurately resolve the rapidly oscillating dipole dynamics, confirming that the numerical stability of our implementation is robust even under extreme field conditions.

\subsection{Eigenvalue problem}
\label{sec:eigensolver}

We obtain bound states and the initial ground state by solving the generalised eigenvalue problem \eqref{eq:gen_ev} using SLEPc's Krylov–Schur method with shift‐and‐invert spectral transformation. The shift $\sigma$ is placed near the expected ground state energy (e.g., $\sigma = -0.5$~a.u. for hydrogen). We treat the problem as a generalised Hermitian eigenproblem. The inner linear solver for $( \mathbf{H} - \sigma\mathbf{M} )^{-1}$ we select automatically based on the number of DOFs. By default, the code uses Flexible Generalized Minimal Residual method (FGMRES) with block Jacobi and incomplete  factorisation on each block; reverse Cuthill–McKee (RCM) ordering is applied \cite{balay2025petsc, hernandez2005slepc}. In addition, a direct LU solver (MUMPS or cuSPARSE, depending on the matrix type) can be enabled via command line; this overrides the iterative selection above. To compute bulk eigenvalues efficiently, the Krylov–Schur subspace dimension 
\texttt{ncv} and the maximum projected dimension \texttt{mpd} scale 
dynamically with the requested number of eigenvalues $n_{\mathrm{ev}}$. 
This optimisation balances spectral resolution against computational overhead, 
ensuring robust convergence stability while preventing the dense projected 
problem from becoming a serial bottleneck. This heuristic ensures that near‐degenerate manifolds are reliably captured while keeping memory usage under control. The convergence tolerance for the eigensolver is set to $\texttt{rtol}=10^{-12}$, with a maximum of $2\,000$ outer iterations. The inner FGMRES solver uses $\texttt{rtol}=10^{-10}$ (or $10^{-8}$ for huge problems) and a divergence tolerance $\texttt{dtol}=10^{3}$ to prevent silently failing solves. We choose these parameters to balance accuracy and performance.

\subsection{Physics: HDF5 Output}
\label{sec:hdf5}
\begin{figure}[htp!]
\centering
\includegraphics[width=0.5\textwidth]{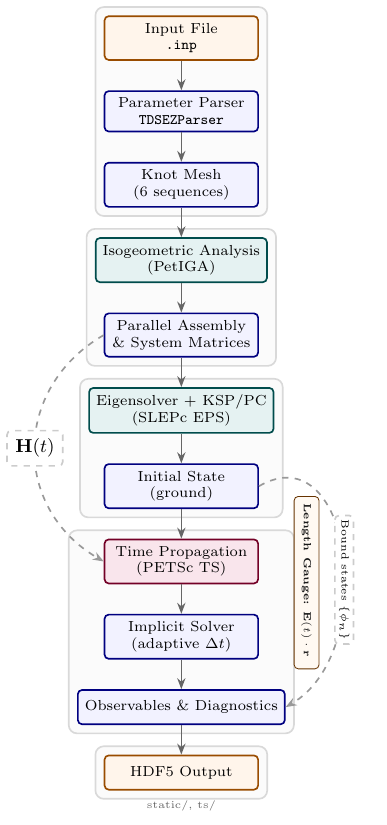}
\caption{Code workflow of TDSE-Z. From input file through geometry/mesh generation, IGA basis construction, parallel assembly of system matrices (mass, BDD kinetic, potential, length-gauge laser, CAP), eigenproblem (Krylov-Schur + shift-invert), initial state preparation, PETSc/TS Crank-Nicolson time propagation, observable computation, automated diagnostics (TRK, dipole-matrix Hermiticity, $[H,L_z]$ commutator), to structured HDF5 output. Dashed arrows indicate run-time feedback loops (length-gauge dipole precompute, bound-state projections). Green = core innovation (BDD operator, IGA, adaptive solvers); Red = time propagation; Yellow = I/O.}
\label{fig:workflow}
\end{figure}

TDSE-Z writes all simulation data to a structured set of HDF5 files using PETSc with collective I/O and time-stepping support. Three files are produced per run, named by prefix and input file:

\begin{enumerate}
\item \texttt{static/EigenData\_\textless input\textgreater.h5} -- Static eigenproblem data (written once after diagonalisation):
\begin{itemize}
\item \texttt{spectrum:} Vector of computed eigenvalues for the requested $n_{\mathrm{ev}}$ states.
\item \texttt{psi\_$i$:} Bound-state wavefunctions (if \texttt{NBoundStatesSave=1}), named by index.
\item \texttt{knots\_x}, \texttt{knots\_y}, \texttt{knots\_z:} Knot vectors per dimension (if wavefunctions saved), enabling exact post-processing reconstruction of the basis.
\end{itemize}

\item \texttt{td/ts\_\textless input\textgreater.h5:} Time-series observables (written every step via PETSc time-stepping):

\begin{itemize}
\item \texttt{populations:} $[t,\; |c_0|^2,\; |c_1|^2,\; \dots,\; |c_{N_{\mathrm{pop}}-1}|^2]$ where $c_n(t)=\langle\phi_n|\Psi(t)\rangle$ are projections onto the bound eigenbasis. 

\item \texttt{dipoles:} Time-dependent data array structured as $[t, \{E_i(t)\}, \{d_i(t)\}, \{a_i(t)\}]$. Here, $E_i(t)$ is the electric field, $d_i(t) = \langle\Psi|i|\Psi\rangle$ is the length-gauge dipole moment, and $a_i(t)$ is the dipole acceleration.

\item \texttt{energy:} The kinetic, potential, laser-interaction, total energy, average inverse mass, and norm are packed as follows: $[t,\; E_{\mathrm{kin}},\; E_{\mathrm{pot}},\; E_{\mathrm{int}},\; E_{\mathrm{tot}},\; \langle m^{-1}\rangle,\; \|\Psi\|^2]$: 

\item \texttt{currents:} Time-dependent array structured as $[t, \langle L_z\rangle, \gamma, \{J_i^{\mathrm{tot}}\}, \{J_i^{\mathrm{intra}}\}, \{J_i^{\mathrm{inter}}\}, \{J_i^{\mathrm{bc}}\}]$. Here, $\langle L_z\rangle$ is the angular momentum expectation value, $\gamma$ is the total phase, and $J_i$ denotes the total, intraband, interband, and bound-continuum current components, respectively.

\item  \texttt{autocorrelation: } Autocorrelation function: $[t,\; \mathrm{Re}\langle\Psi(0)|\Psi(t)\rangle,\; \mathrm{Im}\langle\Psi(0)|\Psi(t)\rangle]$ via $M$-weighted overlap.
\end{itemize}

\item \texttt{td/wfs\_\textless input\textgreater.h5:} Wavefunction snapshots (written every \texttt{OutputStrideWFS} steps). This enables movie generation of wavepacket and post-hoc observable computation.

\end{enumerate}
where the coordinate indices span $i \in \{x\}$ in 1D, $i \in \{x,y\}$ in 2D, and $i \in \{x,y,z\}$ in 3D. Every dataset processed through PETSc's HDF5 time-stepping API appends an implicit step index, generating clean, sequential multi-dimensional arrays instantly compatible with parallel file parsers in Python (\texttt{h5py}), MATLAB, or C++. Fully collective MPI I/O is enforced throughout, guaranteeing scalable throughput and preventing write serialisation bottlenecks on high-performance parallel filesystems. The complete sequence of execution stages—from initial parameter parsing down to this parallel I/O layer—is schematically outlined in the code workflow of Fig.~\ref{fig:workflow}.

\section{Scalability at production scale}
\label{sec:scaling}

We demonstrate the strong scaling of the framework using a 3D hydrogen potential. We utilised $97$ B-splines per axis with degree $p=3$ and the hydrogenic knot sequence over the domain $[-25,\,25]^3$~a.u. Tests were performed on a homogeneous AMD Zen 4 cluster with a 32-MPI-rank-per-node topology. The ground-state energy for this grid is $E_0 = -0.499995$~a.u.\ (without regularisation). Four end-to-end phases are timed: (i) Hamiltonian matrix assembly, (ii) ground-state eigensolve, (iii) Crank–Nicolson propagation over 100 time-steps, and (iv) HDF5 I/O. The scaling data up to 256 cores is shown in Fig. \ref{fig:scaling} and Table~\ref{tab:scaling-zen4}. From the data, three observations emerge:

\begin{figure}[ht!]
\centering
\includegraphics[width=0.46\textwidth]{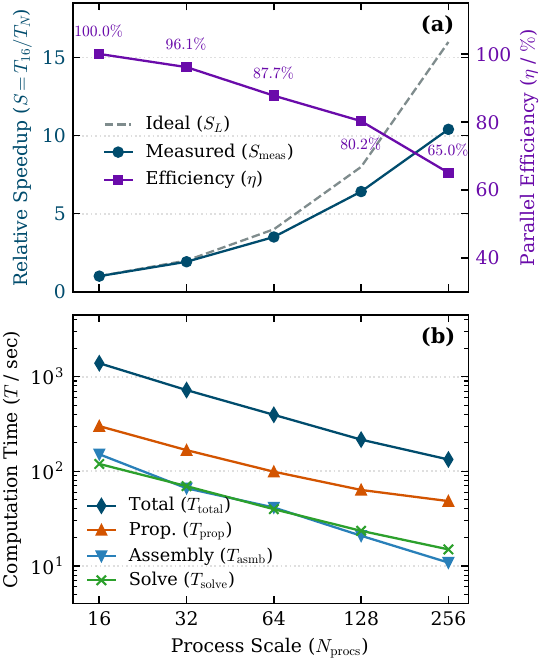}
\caption{Strong scaling on AMD Zen~4. \textbf{(a):}
relative speed-up $S=T_{16}/T_N$ against the number of MPI ranks $N\in\{16,\,32,\,64,\,128,\,256\}$, with ideal $1/N$ scaling overlaid; the parallel efficiency $\eta = S/(N/16)$ is annotated at each rank and
falls from 96\% (32 ranks) to 65\% (256 ranks).
\textbf{(b):} wall time in log scale, decomposed into matrix assembly
(red), ground-state solve, Crank-Nicolson propagation, HDF5 output, and Total time.}
\label{fig:scaling}
\end{figure}

\begin{table}[htbp]
\centering
\caption{Strong-scaling of TDSE-Z on AMD Zen 4. The test problem is 3D Hydrogen driven by an 800 nm laser pulse ($A=0.053$~a.u., $\omega=0.057$~a.u., $T=1$~a.u., $\Delta t=0.01$). $912\,673$ DOFs across 16--256 ranks.
\label{tab:scaling-zen4}}
\begin{tabular}{r r r r r r r}
\toprule
$N_{\rm proc}$ & Assembly & Solve & Propagation & I/O & Total & $\eta_{\rm par}$ (\%) \\
\midrule
 16 & 151.4 & 119.6 & 300.2 & 0.030 & 1387.0 & 100.0 \\
 32 & 66.2  & 69.7  & 167.3 & 0.031 & 721.5  & 96.1 \\
 64 & 41.5  & 39.8  & 99.0  & 0.042 & 395.3  & 87.7 \\
 128 & 20.8 & 23.5  & 63.5  & 0.046 & 216.1  & 80.2 \\
 256 & 10.9 & 15.0  & 48.4  & 0.031 & 133.3  & 65.0 \\
\bottomrule
\end{tabular}
\end{table}

\begin{enumerate}
\item \textbf{Assembly scales near-ideally.}
Assembly time decreases from 151~s (16 ranks) to 10.9~s (256 ranks), achieving $87\%$ efficiency. This scaling reflects the locality of the Galerkin weak form, where work is dominated by element-wise contractions with no global communication. The mild loss of efficiency at high core counts is primarily due to load imbalance induced by the non-uniform hydrogenic knot distribution.

\item \textbf{Propagation is communication- and granularity-limited.}
Crank–Nicolson propagation improves from 300.2~s to 48.4~s (a $6.2\times$ speed-up), but efficiency decreases at higher rank counts due to: (i) increased global synchronisation in Krylov/MUMPS solves (reductions and halo exchanges), and (ii) insufficient problem size per rank, where each rank owns too few DOFs to amortise communication costs.

\item \textbf{I/O is negligible.}
HDF5 output remains below 50~ms across all runs and does not contribute to scaling behaviour.
\end{enumerate}

Overall, efficiency remains above $80\%$ up to 128 ranks, decreasing to $65\%$ at 256 ranks. This performance profile marks the onset of the strong-scaling limit for the $\sim 10^6$-DOF regime, driven by the combined effects of MPI communication overhead, global reductions in the eigensolver, and reduced computation-per-rank associated with over-decomposition.

\section{Results and Validation}
\label{sec:results}
We organise the validation in two tiers: (i) a benchmark proving the central scientific claim—the Quesne semi-confined harmonic oscillator with $m^*(x)=(1+x/a)^{-1}$, an analytical position-dependent-mass (PDM) test case that, until now, has not been matched with machine-precision accuracy by a public TDSE solver; and (ii) a corollary catalogue, where setting $m^*\equiv 1$ reduces the scheme to the canonical atomic TDSE. Verification using harmonic oscillator, hydrogen, and driven-oscillator benchmarks establishes that the BDD machinery introduces no penalty in the constant-mass limit. The following sections present the HHG physics demonstration, the DQW heterostructure, and comparison to existing codes.

\subsection{The Quesne model}
We benchmark the code against an analytical harmonic oscillator model proposed by Quesne \cite{quesne2022generalized}. This model uses a point canonical transformation (PCT) to construct a PDM and potential pair sharing the exact energy spectrum of a standard harmonic oscillator. For the primary case ($m_0=1$), the mass is defined as
\begin{equation}
 M(x) = \left( 1 + \frac{x}{a} \right)^{-1}, \quad x > -a,
 \label{eq:mass}
\end{equation}
which establishes a hard wall at $x = -a$, confining the particle to the domain $x \in (-a, +\infty)$. The theoretical effective potential derived in \cite{quesne2022generalized} relies on a convention where $\hbar = 2m_0 = 1$. Consequently, the kinetic operator in the reference lacks the $1/2$ coefficient present in standard atomic units. The derived potential (Eq.~(20) of Ref.~\cite{quesne2022generalized}) is:
\begin{equation}
 V_{\text{paper}}(x) = \frac{a \omega^2}{4(x+a)} \left( x + a - \frac{\alpha}{a\omega} \right)^2,
 \label{eq:vpaper_general}
\end{equation}
where setting $\alpha = a^2\omega$ recovers the original Jafarov–Van-der-Jeugt (JV) semi-confined harmonic oscillator:
\begin{equation}
 V_{\text{paper}}(x) = \frac{a\omega^2 x^2}{4(x+a)}.
 \label{eq:vpaper_jv}
\end{equation}
Because the numerical solver incorporates the $1/2$ kinetic scaling factor, directly applying $V_{\text{paper}}$ would cause the spectrum to be scaled by $1/2$. To obtain the correct eigenvalues, we rescale both the potential and the energies:
\begin{equation}
 V_{\text{code}}(x) = \frac{1}{2} V_{\text{paper}}(x), \quad E_n^{\text{code}} = \frac{1}{2}\omega\left(n + \frac{1}{2}\right).
 \label{eq:rescaling}
\end{equation}
For the benchmark parameters $a=2$ and $\omega=1$, we executed the benchmark using degree $p=7$ B-splines over 8,000 grid elements with 10-point Gauss–Legendre quadrature. The computational domain was set to $[-2.0, 50.0]$~a.u.\ to capture the singularity boundary. As shown in Table~\ref{tab:eigenvalues}, the numerical eigenvalues match the rescaled analytical solution to within double-precision machine epsilon.

\begin{table}[hbt!]
 \centering
    \caption{Eigenvalue convergence against the analytical Quesne benchmark ($a=2,\ \omega=1$)}
 \label{tab:eigenvalues}
 \begin{tabular}{@{}cccc@{}}
 \toprule
 \textbf{$n$} & \textbf{$E_n^{\mathrm{exact}}$} & \textbf{$E_n^{\mathrm{numeric}}$} & \textbf{Solver $\text{rtol}$} \\
 \midrule
 0 & 0.2500 & 0.2500 & $2.17 \times 10^{-14}$ \\
 1 & 0.7500 & 0.7500 & $1.11 \times 10^{-14}$ \\
 2 & 1.2500 & 1.2500 & $9.21 \times 10^{-15}$ \\
 3 & 1.7500 & 1.7500 & $8.15 \times 10^{-15}$ \\
 4 & 2.2500 & 2.2500 & $6.95 \times 10^{-15}$ \\
 5 & 2.7500 & 2.7500 & $6.90 \times 10^{-15}$ \\
 6 & 3.2500 & 3.2500 & $6.63 \times 10^{-15}$ \\
 7 & 3.7500 & 3.7500 & $6.53 \times 10^{-15}$ \\
 8 & 4.2500 & 4.2500 & $6.01 \times 10^{-15}$ \\
 9 & 4.7500 & 4.7500 & $5.59 \times 10^{-15}$ \\
 \bottomrule
 \end{tabular}
\end{table}

The successful extraction of the rescaled harmonic oscillator spectrum establishes the accuracy of the B-spline BenDaniel–Duke implementation. This validation confirms both the continuous probability flux conservation and the algebraic scaling required for position-dependent mass frameworks.

\subsection{Double quantum well structure}
\label{sec:dqw}

Having validated the BDD framework on analytical PDM models and constant-mass benchmarks, we now apply it to a technologically relevant setting: a symmetric $\text{GaAs/Al}_{0.3}\text{Ga}_{0.7}\text{As}$ double quantum well (DQW). This system exhibits discontinuous mass and potential profiles at material interfaces, providing a stringent stress test of the BDD weak-form assembly. The symmetric DQW consists of two GaAs quantum wells of width $L_w = 8$~nm, separated by a central $\text{Al}_{0.3}\text{Ga}_{0.7}\text{As}$ barrier of width $w_b = 4$~nm and surrounded by thick outer barriers \cite{adachi1994gaas, levinshtein1997handbook}. The conduction-band profile and effective mass are piecewise constant:
\begin{align}
V(x) &= \begin{cases}
V_0, & |x| < w_b/2 \quad \text{(central barrier)}, \\
0, & w_b/2 < |x| < w_b/2 + L_w \quad \text{(wells)}, \\
V_0, & |x| > w_b/2 + L_w \quad \text{(outer barriers)},
\end{cases} \\
m^*(x) &= \begin{cases}
m^*_b = 0.092\,m_e, & \text{barrier regions}, \\
m^*_w = 0.067\,m_e, & \text{well regions}.
\end{cases}
\end{align}
The potential height is $V_0 \approx 233.7$~meV.

\subsubsection{Tunnelling splitting}
For $w_b = 4$~nm, the production $N_{\rm splines}=15\,000$ sweep gives a ground-state doublet energy splitting $\Delta E_{\text{BDD}} \equiv E_1 - E_0 = 2.8615$~meV ($E_0 = 40.6084$~meV, $E_1 = 43.4699$~meV). We validate this against semiclassical WKB theory \cite{landau1958quantum, griffiths2018introduction}. For square-barrier heterostructures, the leading-order WKB prefactor is of order unity \cite{harrison2016quantum,bastard1989wave}. The imaginary momentum $\kappa$ within the central barrier is constant:
\begin{equation}
\kappa = \sqrt{2m^*_b (V_0 - E_0)}.
\end{equation}
The action integral $S = \kappa w_b \simeq 2.733$ yields a suppression factor $e^{-S} \simeq 0.0650$. Using the effective well frequency $\hbar\omega_{\text{eff}} = 43.50$~meV for an isolated 8~nm GaAs well gives the WKB estimate $\Delta E_{\text{WKB}} = \hbar\omega_{\text{eff}} e^{-S} \simeq 2.83$~meV, within about $1.2\%$ of $\Delta E_{\text{BDD}}$. This confirms that the high-resolution B-spline discretisation captures exponential attenuation while enforcing BDD flux-matching at sharp interfaces.

\begin{figure}[t]
\centering
\includegraphics[width=\columnwidth]{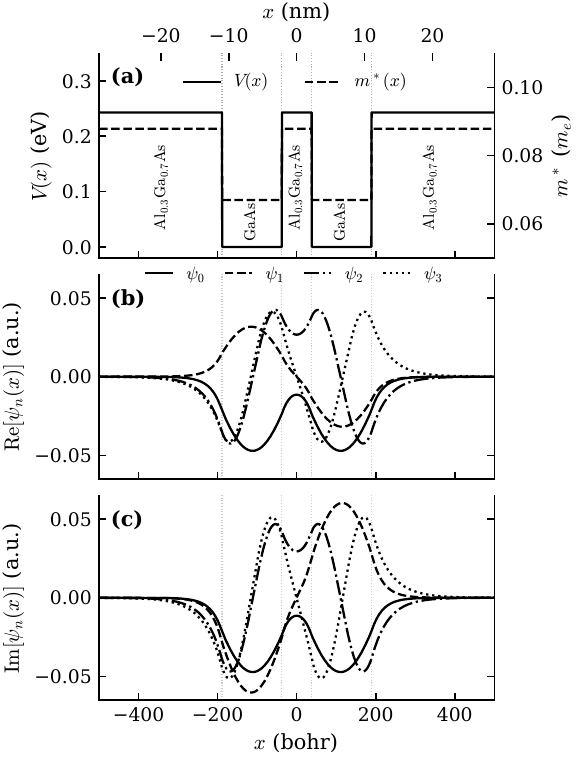}
\caption{(a) Conduction-band potential profile $V(x)$ (solid curve, left axis, in eV) and position-dependent effective mass $m^*(x)$ (dashed curve, right axis, in units of $m_e$) for the symmetric $\text{GaAs/Al}_{0.3}\text{Ga}_{0.7}\text{As}$ double quantum well structure. The configuration defines an inner $4$~nm central barrier bounded by two symmetric $8$~nm quantum wells, with step heterojunctions located at $x = \pm 37.79$~bohr ($\pm 2$~nm) and $\pm 188.97$~bohr ($\pm 10$~nm). (b) Real and (c) imaginary components of the first four stationary bound-state eigenfunctions ($\psi_0$ to $\psi_3$). The vanishing imaginary channels reflect the pure real-symmetric nature of the static Hamiltonian. The BenDaniel-Duke operator handles mass discontinuities by sampling $1/m^*(x)$ at Gauss-Legendre quadrature points, ensuring matching and continuous probability flux across all boundaries.}
\label{fig:dqw}
\end{figure}

\subsubsection{Barrier-width sweep}
The most stringent WKB test is the exponential scaling of $\Delta E$ with $w_b$. We compute $\Delta E$ for $w_b \in \{2, 4, 6, 8, 10\}$~nm; results are summarised in Table~\ref{tab:dqw-sweep}. A log-linear fit yields a decay length $\xi_{\rm fit} = 1.469$~nm, which agrees with the theoretical prediction $\xi_{\rm theory} = 1.463$~nm to within $0.4\%$.

\begin{table}[htbp]
\centering
\caption{Computed DQW energies for varying barrier width $w_b$. Convergence verified at $15\,007$ B-splines (shifts $<0.1\,\%$).}
\label{tab:dqw-sweep}
\begin{tabular}{cccc}
\toprule
$w_b$ (nm) & $E_0$ (meV) & $E_1$ (meV) & $\Delta E$ (meV) \\
\midrule
2 & 36.21 & 47.35 & 11.14 \\
4 & 40.61 & 43.47 & 2.86 \\
6 & 41.68 & 42.42 & 0.73 \\
8 & 41.96 & 42.14 & 0.19 \\
10 & 42.02 & 42.07 & 0.048 \\
\bottomrule
\end{tabular}
\end{table}
For solid-state heterostructures, the solver strictly enforces probability flux continuity across sharp material interfaces; computed tunnel splittings in a $\text{GaAs/Al}_{0.3}\text{Ga}_{0.7}\text{As}$ double quantum well follow the semiclassical WKB exponential decay over two orders of magnitude, with a fitted decay length within $0.4\%$ of theory.

\begin{figure}[htbp]
\centering
\includegraphics[width=0.4\textwidth]{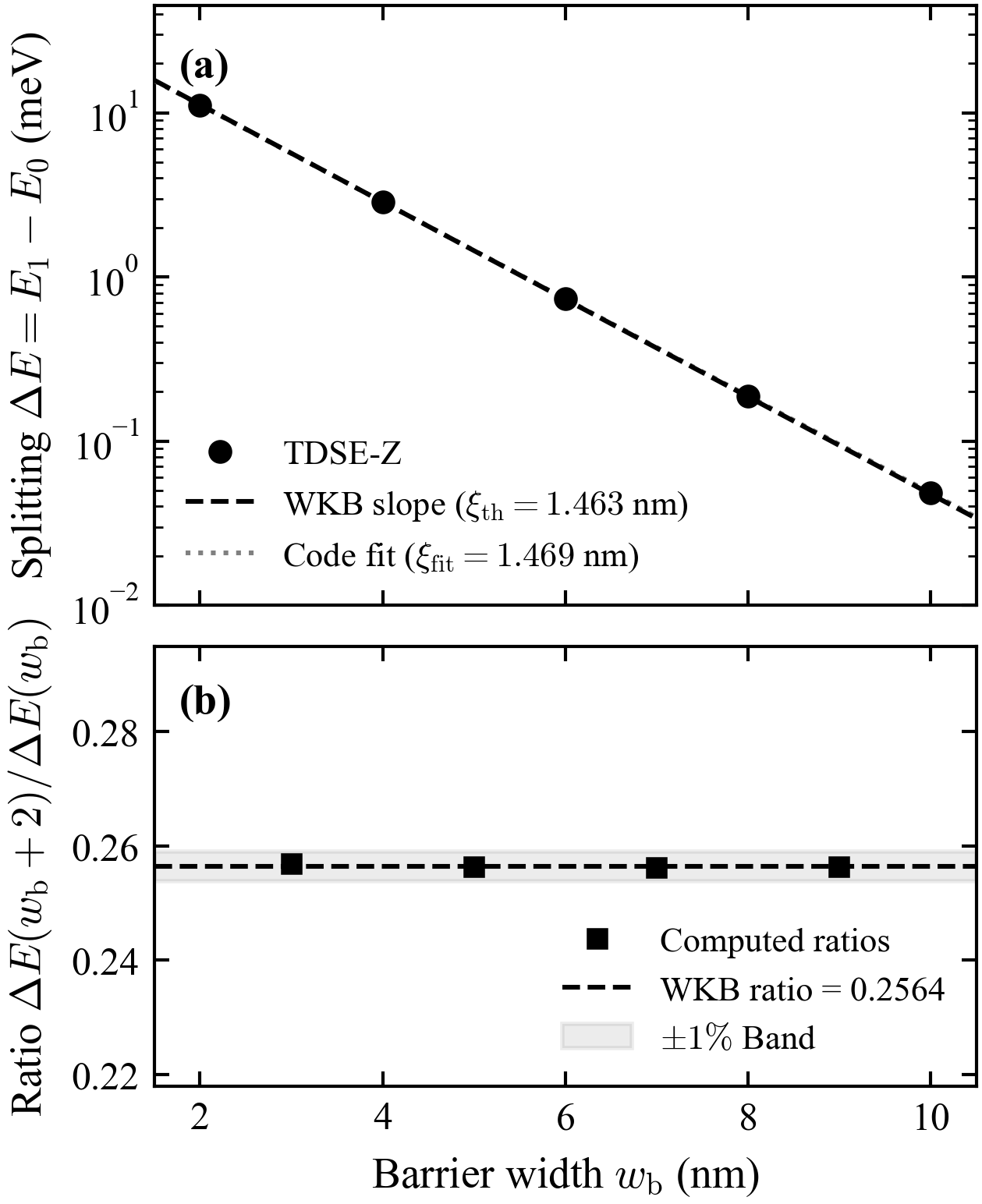}
\caption{Computed ground-state energy $E_0$, first excited state $E_1$, and splitting $\Delta E$ for the symmetric DQW as a function of barrier width $w_b$. Basis convergence verified by doubling the B-spline count; all energies shift by less than $0.1\,\%$.}
\label{fig:dqw-wkb-validation}
\end{figure}

\subsection{Molecular HHG and time-frequency dynamics}
\begin{widetext}
\begin{figure*}[!htbp]
 \centering
 \includegraphics[width=0.8\linewidth]{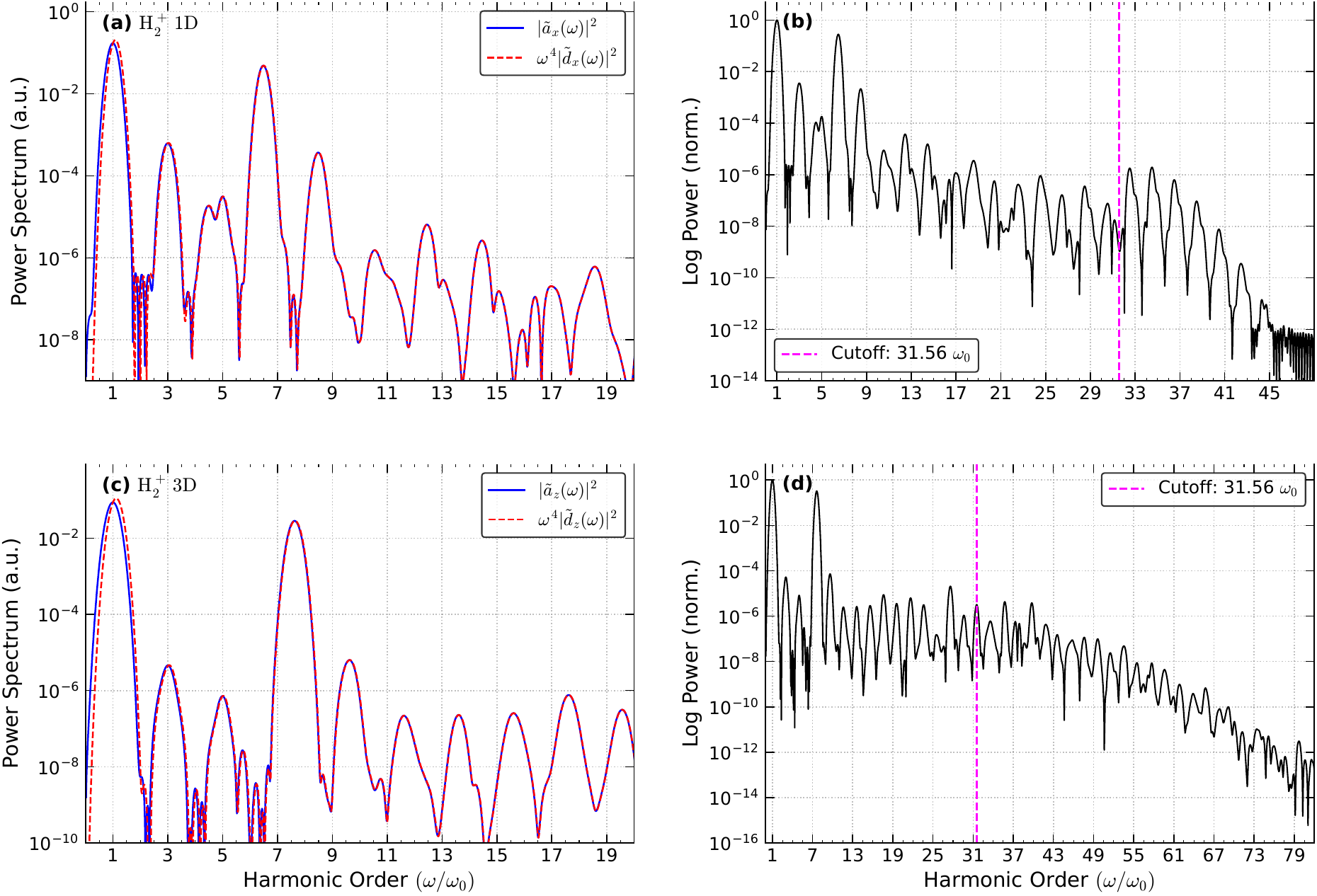}
    \caption{HHG spectra of $\mathrm{H}_2^+$ computed in one dimension (1D, top row) and three dimensions (3D, bottom row). Panels (\textbf{a}) and (\textbf{c}) display the validation of gauge invariance up to the 20th harmonic order, comparing the power spectra obtained via the acceleration gauge ($|\tilde{a}_{x,z}(\omega)|^2$, solid blue lines) and the frequency-scaled length gauge ($\omega^4 |\tilde{d}_{x,z}(\omega)|^2$, dashed red lines). Panels (\textbf{b}) and (\textbf{d}) show the normalized log-power spectra evaluated over an extended frequency range to highlight the plateau and cutoff configurations. The vertical dashed lines mark the semiclassical three-step model cutoff. In stark contrast to the 1D case, the 3D spectrum exhibits a dramatic  yield and shape difference of the high harmonics, especially in the region beyond this classical boundary, a consequence of dimensional wave-packet spreading DoFs.}
 \label{fig:hhg_h2p}
\end{figure*}
\end{widetext}

\begin{figure}[htp!]
 \centering
 \includegraphics[width=1.01\linewidth]{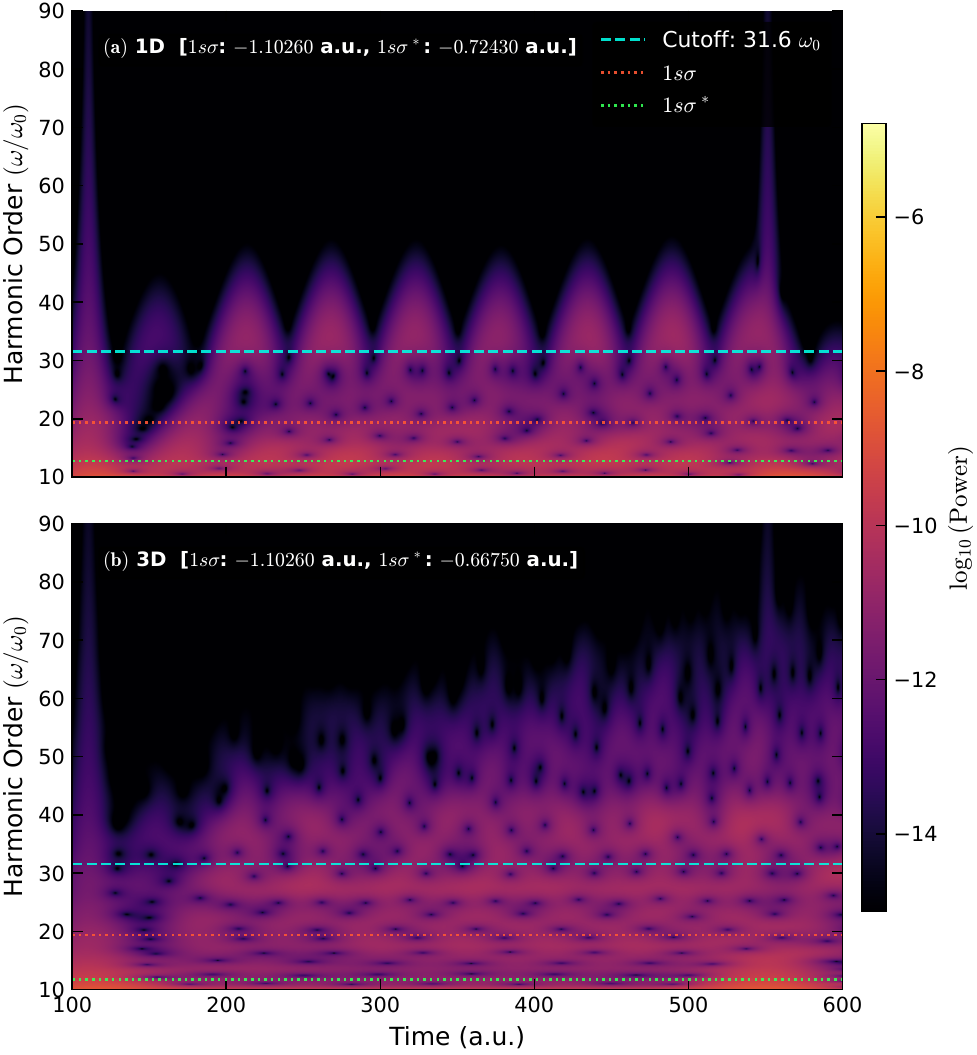}
    \caption{Time-frequency analysis of molecular HHG. Continuous wavelet transform (CWT) spectrograms of the high-harmonic emission for (a) the 1D model evaluated using $W = 11$ and (b) the 3D \ce{H2+} model evaluated using $W = 17$ to account for higher continuum noise. The colour map represents the logarithmic emission power. The horizontal dashed green line marks the semiclassical three-step cutoff at $E_{\text{cutoff}} \approx 31.56\,\omega_0$. A continuous, steady-state emission band is visible in both panels at $6.6421\,\omega_0$ in (a) and $7.6421\,\omega_0$ in (b), corresponding to field-driven bound-bound Rabi cycling between the $1s\sigma$ and $1s\sigma^*$ charge-resonance states. The 1D framework (a) exhibits a high-contrast sub-cycle quantum-paths contributions, sustained by the artificial phase-amplitude balance of strictly collinear short and long trajectories. In contrast, the 3D spectrogram (b) shows more complex quantum-paths contributions due to transverse wave-packet expansion in the extra spatial dimensions, which include multi-returns contributions from non-subsequent half cycles .}
 \label{fig:wavelet_h2p}
\end{figure}

To elucidate the role of dimensionality in strong-field molecular interactions, we compare HHG from a 1D model of \ce{H2+} against a full 3D treatment under equivalent laser parameters. The nuclei are clamped at internuclear distance $R = 2.0\text{ a.u.}$ within the Born–Oppenheimer approximation. The 3D two-centre potential is $V_{\text{3D}}(\mathbf{r}) = -1/\sqrt{x^2+y^2+(z+R/2)^2} - 1/\sqrt{x^2+y^2+(z-R/2)^2}$. The 1D model, restricted to the laser polarisation axis, employs the soft-core potential $V_{\text{1D}}(x) = -1/\sqrt{(x+R/2)^2 + \epsilon} - 1/\sqrt{(x-R/2)^2 + \epsilon}$, where $\epsilon = 1.472\text{ a.u.}$ regularises the 1D potential and is calibrated to reproduce the 3D ionisation potential, yielding $I_p^{\text{1D}} \approx I_p^{\text{3D}} \approx 1.1026\text{ a.u.}$ This ensures energetic equivalence while making spatial dimensionality the sole variable. We use 30,007 degree-7 B-splines over the interval $[-4000, 4000]\text{ a.u.}$ for the 1D calculations and 161 degree-3 B-splines per axis over $[-80, 80]\text{ a.u.}$ for 3D. A hydrogenic knot sequence is adopted for both simulations, reproducing the ground-state energy $E_0 \approx -1.1026~\mathrm{a.u.}$ in both cases. The first ungerade excited state differs between the calibrated 1D and full 3D models: $E_1^{\mathrm{1D}} \approx -0.72430~\mathrm{a.u.}$ and $E_1^{\mathrm{3D}} \approx -0.6675~\mathrm{a.u.}$, consistent with the labels in Fig.~\ref{fig:wavelet_h2p} and with the 3D reference values of Ishikawa \textit{et al.}~\cite{ishikawa2008solving}. Both systems are irradiated with a 6-cycle trapezoidal pulse ($\omega_0 = 0.057\text{ a.u.}$, $I = 1 \times 10^{14}\text{ W/cm}^2$) linearly polarised along the molecular axis~\cite{labeye2018optimal}. Figures~\ref{fig:hhg_h2p}(a) and~\ref{fig:hhg_h2p}(c) demonstrate excellent consistency between the acceleration $|\tilde{a}(\omega)|^2$ and frequency-scaled length-gauge $\omega^4|\tilde{d}(\omega)|^2$ spectra, confirming numerical convergence.

\begin{figure}
    \centering
    \includegraphics[width=0.9\linewidth]{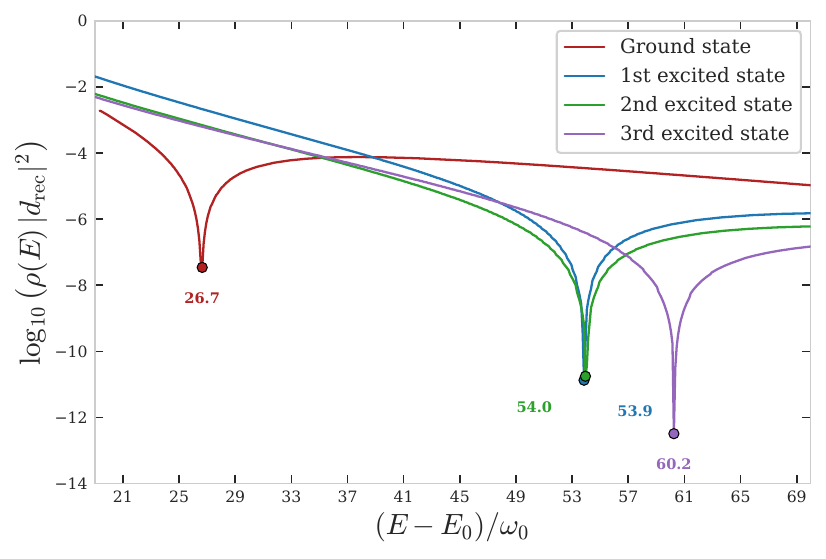}
    
    \caption{Logarithm of the differential recombination probability, $\log_{10}(\rho(E) |d_{\text{rec}}|^2)$, as a function of the scaled photoelectron energy $(E - E_0)/\omega_0$ for the ground state (red) and the first (blue), second (green), and third (purple) excited states of 1D $\mathrm{H}_2^+$.  $d_{\text{rec}} (E_i) = \braket{\psi_i|\hat{D}_x|\psi_{E_i}}$ is the dipole recombination and $\rho(E_i) = (E_{i+1} - E_i)^{-1}$ is the density of continuum states. The sharp destructive interference minima are indicated by the circular markers, located at scaled energies of $26.7$ for the ground state, $53.9$ and $54.0$ for the first and second excited states, and $60.2$ for the third excited state.}
    
    \label{fig:dipole_rec}
\end{figure}

As observed in Fig.~\ref{fig:hhg_h2p}, spatial dimensionality profoundly modifies the low-order harmonic yield and structure \cite{chirilua2006strong, lewenstein1994theory}. This is a well-known effect observed from 1D to 3D HHG modelling, where absolute harmonic yields, among other properties, are dependent upon this dimensionality and lead, for instance, to a higher estimated contribution of the long trajectory in the plateau region for the 1D model. This is because in 1D the transversal spreading of the electronic wavepacket for a given trajectory is absent, preventing the consideration of orbital symmetry effects such as Cooper minima, fully described Coulomb focusing close to recombination, and harmonic yield and polarisation angular dependence. However structural fixture can be capture in both. A good example is two-centres interferences signature for molecular HHG.

This central feature of the molecular response manifests into a structural interference minimum in the dipole recombination cross-section, shown in Fig.\ref{fig:dipole_rec} using our 1D reference framework. For the ground state ($1s\sigma_g$) at $R = 2.0\text{ a.u.}$, the static two-centre interference condition $k R \cos\theta = \pi$—originally established by Lein \textit{et al.} \cite{lein2002interference, lein2002role} predicts a minimum at $21.6\ \omega_0$. However, our numerical analysis identifies this minimum at a scaled energy of $26.7$ [Fig.\ref{fig:dipole_rec}]. This blue-shift is a direct signature of dynamic orbital contraction and laser-dressed potential effects, which effectively shorten the perceived internuclear distance. Interestingly, the excited states exhibit minima at significantly higher energies—$53.9$ and $54.0$ for the first and second excited states, and $60.2$ for the third excited state—reflecting their more complex nodal topologies. The 1D spectrum [Fig.\ref{fig:hhg_h2p}(b)] clearly preserves the ground-state structural signature as a deep suppression near the 27th harmonic. In contrast, while the full 3D treatment [Fig.\ref{fig:hhg_h2p}(d)] involves prohibitive computational costs for direct orbital-resolved recombination extraction, the resulting harmonic spectra demonstrate how transverse momentum components and non-collinear return paths modify the overall high-frequency emission plateau.

The high-energy spectral region in Fig.~\ref{fig:hhg_h2p} further exposes the inadequacy of reduced-dimensional models. If the semiclassical cutoff \cite{l1993high}, $E_{\text{cutoff}} = I_p + 3.17U_p \approx 31.56 \omega_0$, is identical in both frameworks, the yield of HHG beyond the cut-off is overestimated in 1D [Fig.~\ref{fig:hhg_h2p}(b)] compared to 3D result [Fig.~\ref{fig:hhg_h2p}(d)]. This is mainly due to overestimated recollision probability being artificially high in 1D as the electronic wavepacket is forced to move along one polarisation axis; the Coulomb focusing is also artificially described in 1D with the scattering dynamics being not correctly described.

To resolve the sub-cycle origins of these disparities, we employ a continuous wavelet transform (CWT) with resolution parameters $W = 11$ for the 1D case and $W = 17$ for the 3D case to account for the higher continuum noise in 3D [Fig. \ref{fig:wavelet_h2p}]. Across both dimensionalities, a continuous high-intensity horizontal band is observed at transition energies of $6.6421\ \omega_0$ in 1D and $7.6421\ \omega_0$ in 3D, corresponding to the resonant $1s\sigma \leftrightarrow 1s\sigma^*$ transition; this signifies continuous Rabi flopping. The 1D wavelet map [Fig.~\ref{fig:wavelet_h2p}(a)] reveals a characteristic interference grid sustained by the comparable amplitudes of short and long trajectories. In the 3D map [Fig.~\ref{fig:wavelet_h2p}(b)], this grid is suppressed, providing time-domain evidence for the dramatic spatial spreading: long-trajectory components fail to return to the molecular core with sufficient density to generate measurable sub-cycle interference. Beyond the classical cutoff, the 3D wavelet map reveals no localised vertical emission columns above the $31.56 \omega_0$ boundary. This implies that the 3D extended plateau originates from temporally diffuse, non-local multi-centre recollision pathways rather than high-intensity attosecond bursts. Reduced-dimensional models therefore systematically overestimate HHG coherence; full-dimensional modelling is indispensable for correctly interpreting the high-energy molecular response. This plays an important role for understanding HHG in heterostructure systems.

\section{Conclusion}
\label{sec:concl}

We have introduced \textsc{TDSE-Z}, a unified high-performance computational framework that bridges a critical gap in strong-field physics: the accurate simulation of quantum dynamics across both constant-mass atomic/molecular systems and spatially varying effective-mass semiconductor heterostructures. At its core, the framework employs a rigorous weak-form Galerkin discretisation of the Hermitian BenDaniel–Duke operator on geometry-adapted B-spline meshes. Through this approach, we achieve a significant computational milestone: machine-precision agreement with the analytical Quesne position-dependent-mass benchmark—an exacting test that has eluded public TDSE implementations to date.

This mathematical rigour translates directly into physical fidelity across disparate regimes. In the constant-mass limit, the framework seamlessly reproduces canonical atomic TDSE catalogues, including harmonic oscillator eigenvalues, hydrogenic Rydberg series, and analytical Rabi dynamics. For solid-state heterostructures, the solver enforces probability flux continuity across sharp material interfaces; computed tunnel splittings in a $\text{GaAs/Al}_{0.3}\text{Ga}_{0.7}\text{As}$ double quantum well follow the semiclassical WKB exponential decay over two orders of magnitude, with a fitted decay length within $0.4\%$ of theory. This geometry is directly relevant to recent solid-state HHG experiments \cite{ghimire2011observation, liu2017high, vampa2015semiclassical}. Using GaAs well parameters ($m^* = 0.067\,m_e$, band gap $E_g = 1.42$~eV) and a mid-infrared driver ($\lambda = 10\,\mu\text{m}$, $\hbar\omega = 0.124$~eV), the Keldysh parameter indicates regimes ranging from multi-photon ($\gamma \approx 2.3$) to tunnelling-dominated ($\gamma \approx 0.5$). This parameter space and the associated dynamics will be the subject of dedicated investigations reported in forthcoming work.

When applied to molecular strong-field dynamics, \textsc{TDSE-Z} reveals a critical physical insight: the 3D high-harmonic spectrum of $\text{H}_2^+$ extends far beyond the classical cutoff, yet the corresponding time-resolved spectrogram reveals a complete absence of discrete attosecond bursts. We demonstrate that this extended plateau is not driven by coherent recollisions, but by temporally diffuse multi-centre lateral scattering. This finding serves as a warning to the attosecond community: reduced-dimensional models systematically overestimate pulse coherence, and integrated spectra alone are insufficient to diagnose attosecond pulse formation.

We built TDSE-Z on PETSc, SLEPc, and PetIGA ecosystems with a demonstrated strong-scaling efficiency, exceeding $80\%$ up to $128$ cores. The code is suitable for heterogeneous computing: the TISE eigensolver is fully GPU-accelerated for rapid initial-state preparation, and the time-propagation engine leverages highly optimised CPU-based MPI parallelism. \textsc{TDSE-Z} is freely available for academic use, providing the AMO, condensed-matter, and computational physics communities with a production-ready platform to explore laser-driven dynamics across diverse spatial and mass scales. The code could also be used to generate the necessary data to train new machine learning models capable of predicting physics and bypassing expensive simulations.

\section*{Software Availability}
The \textsc{TDSE-Z} framework is provided for academic and non-commercial research purposes. The source code, documentation, and benchmark input files are available in the GitHub repository: \url{https://github.com/dahbiz/tdsez}.

\begin{acknowledgments}
Z.D. and A.Z. are grateful to E. Cormier for insightful discussions during the early stages of this work, and to R. Guichard for helpful feedback on the draft. Z.D. also thanks S. Zampini for his invaluable introduction to the PETSc ecosystem during the MHPC program (ICTP/SISSA), which provided the essential computational basis for this implementation. Z.D. and A.Z. acknowledge funding from UK Research and Innovation (UKRI) under the UK government’s Horizon Europe funding guarantee [Grant No. EP/Z000807/1]. 
\end{acknowledgments}

\appendix
\section{Exact 3D benchmark for the resonant sine drive}

To validate the full 3D propagator and optimisation implemented, we consider an isotropic harmonic oscillator (HO) ($(m=\omega=1$), $(\hbar=1$)) in three dimensions driven along the $(x$)-axis by a resonant field
\begin{equation}
  \mathbf{E}(t) = E_0 \sin(t) \hat{\mathbf{x}},
\end{equation}
where $E_0$ is the peak electric field amplitude. The full Hamiltonian is
\begin{equation}
  H(t) = \sum_{j=x,y,z} \left( \frac{p_j^2}{2} + \frac{1}{2} x_j^2 \right) + x E_0 \sin(t).
\end{equation}
The initial state is the 3D ground state ($\epsilon_0=1.5$~a.u). The numerical propagation is performed in the full 3D space without any dimensional reduction. The goal of this validation is to confirm that TDSE-Z propagation is accurate under strong-laser fields.

\subsection{3D HO Analytic Solution}
For a linear potential, the exact analytic solution is known. Because the Hamiltonian is separable, the exact wavefunction is a product of coherent states:
\begin{equation}
  \ket{\Psi(t)}_{\text{exact}} = \ket{\alpha(t)} \otimes \ket{0} \otimes \ket{0},
\end{equation}
where \(\ket{0}\) denotes the 1D ground state and \(\alpha(t)\) is the coherent-state parameter for the driven \(x\)-mode. This product structure is a property of the exact solution; the numerical solver must reproduce it dynamically from the full 3D equations. The coherent-state parameter is derived via the interaction picture:
\begin{equation}
  \alpha(t) = \frac{E_0}{2\sqrt{2}} \left( t e^{-it} - \sin (t) \right),
\end{equation}
with
\begin{equation}
  |\alpha(t)|^2 = \frac{E_0^2}{8} \left( t^2 + \sin^2 (t) - t \sin(2t) \right).
\end{equation}
The expectation value of the total Hamiltonian $(H(t) = H_0 + x E_0 \sin t$) is
\begin{equation}
  \langle H(t) \rangle_{3D} = \langle H_0 \rangle + E_0 \sin t \, \langle x_{\textrm{ex}} \rangle,
\end{equation}
where $\langle H_0 \rangle = \frac{3}{2} + |\alpha|^2$  and $\langle x_{\textrm{ex}} \rangle = \frac{E_0}{2}(t\cos(t) - \sin(t))$. Substitution yields the closed-form expression for the total energy:
\begin{equation} 
\langle H(t) \rangle_{3D}^{\text{exact}} = \frac{3}{2} + \frac{E_0^2}{8} \left( t^2 + 2t\sin(t) \cos (t) - 3\sin^2 (t) \right).
  \label{eq:Etotal}
\end{equation}
The driven-mode coherent state probabilities are 
\begin{equation}
    P_n (t) = \exp(-|\alpha|^2)\frac{|\alpha|^{2n}}{n!}.
\end{equation}
We compare these analytical quantities against the full 3D numerical results.

\subsection{3D HO Numerical Convergence}
To benchmark numerical accuracy and stability, we compare our 3D TDSE-Z solver against exact analytical solutions across varying time steps ($\Delta t$) and field amplitudes ($E_0$). 

\begin{figure}[ht]
    \centering
    \includegraphics[width=0.9\linewidth]{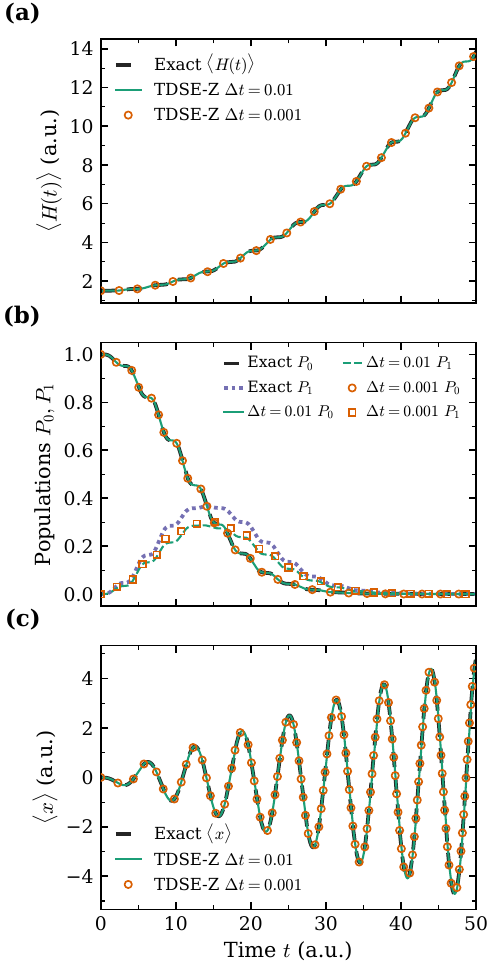}
    \caption{Time evolution of total energy $\langle H(t)\rangle$ (a), shell populations $P_0, P_1$ (b), and position expectation value $\langle x \rangle$ (c) for $E_0 = 0.2$~a.u ($\approx1.4 \times10^{15}$~W/cm$^2$). up to $t = 50$~a.u.}
    \label{fig:ho_comparison}
\end{figure}

Figure~\ref{fig:ho_comparison} displays the time evolution of the total energy $\langle H(t)\rangle$, ground- and first-excited shell populations ($P_0, P_1$), and the position expectation value $\langle x \rangle$ for $E_0 = 0.2$ a.u. Up to $t = 50$ a.u., the finest time step ($\Delta t = 0.001$) exhibits near-perfect agreement with the exact analytical solution. It captures the rapid coherent oscillations without secular drift. 

\begin{figure}[ht!]
    \centering
    \includegraphics[width=0.9\linewidth]{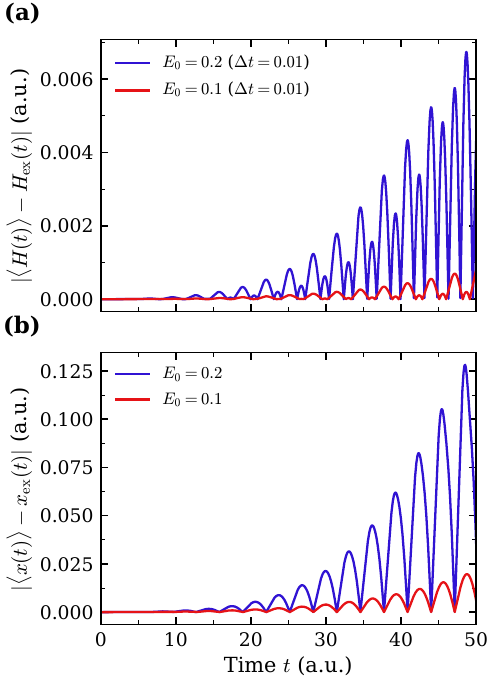}
    \caption{Linear-scale absolute errors in total energy (a) and position observables (b) at $\Delta t = 0.01$ comparing intense ($E_0 = 0.2$)~a.u ($\approx1.4 \times10^{15}$~W/cm$^2$) and ($E_0 = 0.1$)~a.u  ($\approx3.51 \times10^{14}$~W/cm$^2$) laser regimes.}
    \label{fig:ho_error}
\end{figure}

Figure~\ref{fig:ho_error} highlights the critical relation between field strength and temporal discretisation. We evaluate the absolute errors in total energy and position observables at $\Delta t = 0.01$ for $E_0 = 0.2$ and $E_0 = 0.1$. The results provide robust evidence that stronger laser fields demand correspondingly smaller time steps ($\Delta t$) to maintain high numerical fidelity. Specifically, doubling the field amplitude from $E_0 = 0.1$ to $E_0 = 0.2$ results in a substantial amplification of discretisation and truncation errors at coarser resolutions. This underscores the necessity of high-resolution temporal propagation ($\Delta t \to 0.001$) in intense-field regimes to suppress error accumulation and ensure the long-term stability of the 3D TDSE-Z solver.

% \nocite{*}
\bibliographystyle{apsrev4-2}
\bibliography{references}

\end{document}